\documentclass[12pt]{article}
\usepackage{scicite}
\usepackage{times}
\usepackage{graphicx}
\newcounter{lastnote}
\newenvironment{scilastnote}{%
\setcounter{lastnote}{\value{enumiv}}%
\addtocounter{lastnote}{+1}%
\begin{list}%
{\arabic{lastnote}.}
{\setlength{\leftmargin}{.22in}}
{\setlength{\labelsep}{.5em}}}
{\end{list}}
\begin{document}
\baselineskip24pt
\setcounter{page}{1}
\newcommand{\vg}{V_G}
\newcommand{\Vg}{$\vg$}
\newcommand{\vsd}{V_{SD}}
\newcommand{\Vsd}{$\vsd$}
\newcommand{\GVB}{$G\left(\vsd,B\right)$}
\newcommand{\It}{I_{T}}
\newcommand{\dIt}{\partial\It/\partial\vg}
\newcommand{\dItVB}{$\partial \It\left(\vsd,B\right)/\partial\vg$}
\newcommand{\dItVgB}{$\partial \It\left(\vg,B\right)/\partial\vg$}
\newcommand{\micro}[1]{$\mu\mbox{#1}$}
\newcommand{\fig}[1]{Fig.~\ref{#1}}
\newcommand{\figs}[1]{Figs.~\ref{#1}}
\newcommand{\eq}[1]{Eq.~\ref{#1}}
\newcommand{\eqs}[1]{Eqs.~\ref{#1}}
\newcommand{\text}[1]{\tt #1}
\newcommand{\lw}{\tt L}
\newcommand{\uw}{\tt U}
\newcommand{\ul}{\tt U,L}
\newcommand{\bp}{B_+}
\newcommand{\bm}{B_-}
\newcommand{\Bp}{$\bp$}
\newcommand{\Bm}{$\bm$}
\newcommand{\bpm}{B_\pm}
\newcommand{\Bpm}{$\bpm$}
\newcommand{\kfu}{k_{F{\uw}}}
\newcommand{\kfl}{k_{F{\lw}}}
\newcommand{\sub}[1]{${}_{#1}$}
\newcommand{\AkE}{${\cal A}_{k,E}$}

\begin{flushleft}
{\bf\LARGE Spin-charge separation and localization in
one-dimension}\vspace{0.3in}\\%
{\bf\large O.~M. Auslaender,$^{1\dag}$ H. Steinberg,$^{1}$ A.
Yacoby,$^{1\ast}$ Y. Tserkovnyak,$^{2}$ B.~I. Halperin,$^{2}$ K.~W.
Baldwin,$^{3}$ L.~N.
Pfeiffer,$^{3}$ K.~W. West$^{3}$}\vspace{0.3in}\\
\normalsize{$^{1}$Dept. of Condensed Matter Physics, Weizmann
Institute of Science, Rehovot 76100, Israel;\\ $^{2}$Lyman
Laboratory of Physics, Harvard University, Cambridge, MA 02138,
USA;\\ $^{3}$Bell Labs, Lucent Technologies,700 Mountain Avenue,
Murray Hill, NJ 07974, USA.} \vspace{0.3in}\\
\normalsize{$^\dag$Present address: Geballe Laboratory for Advanced
Materials, Stanford University, Stanford, CA 94305, USA.\\
$^\ast$To whom correspondence should be addressed; E-mail:
amir.yacoby@weizmann.ac.il.}\end{flushleft}
{\bf
We report on measurements of quantum many-body modes in ballistic
wires and their dependence on Coulomb interactions, obtained from
tunneling between two parallel wires in a GaAs/AlGaAs
heterostructure while varying electron density. We observe two spin
modes and one charge mode of the coupled wires, and map the
dispersion velocities of the modes down to a critical density, at
which spontaneous localization is observed. Theoretical calculations
of the charge velocity agree well with the data, although they also
predict an additional charge mode that is not observed. The measured
spin velocity is found to be smaller than theoretically predicted.
\\
}%

Coulomb interactions have a profound effect on the behavior of
electrons. The low energy properties of interacting electronic
systems are described by elementary excitations, which interact with
each other only weakly. In two and three-dimensional disordered
metals they are dubbed quasi-particles \cite{Nozieres99}, as they
bear a strong resemblance to free electrons \cite{Altshuler85},
which are fermions carrying both charge and spin. However, the
elementary excitations in one-dimensional (1D) metals, known as
Luttinger-liquids \cite{Tomonaga50,Luttinger63}, are utterly
different. Instead, each is collective, highly correlated and
carries either spin or charge.

We determine the dispersions of the elementary excitations in
one-dimension by measuring the tunneling current, $\It$, across an
extended junction between two long ballistic parallel wires in a
GaAs/AlGaAs heterostructure created by cleaved edge overgrowth (CEO)
\cite{Ophir02}. In this geometry tunneling conserves both energy and
momentum. Each tunneling event creates an electron-hole pair with
total momentum $\hbar k=eBd\equiv\hbar q_B$ and total energy
$E=\left|e\vsd\right|$, in which $2\pi\hbar$ is Plank's constant,
$-e$ is the electron charge, $B$ is magnetic field applied
perpendicular to the plane of the wires, $d$ is the distance between
their centers and \Vsd\ is the voltage-bias between them \cite{SOM}.

The rate of tunneling between the wires depends on the ease of
adding an electron to one wire and a hole to the other, determined
by the electron-hole spectral function, \AkE. For weak inter-wire
interactions, \AkE\ is a convolution of the individual particle
spectral functions, which encode the overlap of electrons (or holes)
with the many-body modes of the coupled-wires. Near $\vsd=0$, in the
limit of temperature $T\rightarrow0$, tunneling is appreciable only
if $\left|q_B\right|=\left|\kfu\pm\kfl\right|$, allowing exchange of
electrons between the Fermi-points, $k_{Fi}=\pi n_i/2$, where $n_i$
is electron density in sub-band $i$, while $i=\uw,\lw$ stands for
sub-bands in the upper, lower wires. At finite energies,
interactions broaden the peaks of the individual particle spectral
functions, in particular giving them a distribution of momenta. In
spite of this, at $E=0$, \AkE\ is sharply peaked at
$k=\left|\kfu\pm\kfl\right|$ for homogeneous wires
\cite{Carpentier02}. Thus, as long as momentum is conserved in the
wires and in the tunnel junction, tunneling near $\vsd=0$ is
enhanced at the same $B\ge0$ values as without interactions:
\begin{equation}\label{Bpm}
\bpm=\frac{\hbar}{ed}\left|\kfu\pm\kfl\right|.
\end{equation}%
For inhomogeneous wires, the $\vsd=0$ line-shape of the spectral
function encodes information on the low energy momentum distribution
of the many-body states \cite{Tserkovnyak03}.

\begin{figure}[ptb]
\begin{center}\includegraphics[width=6in]{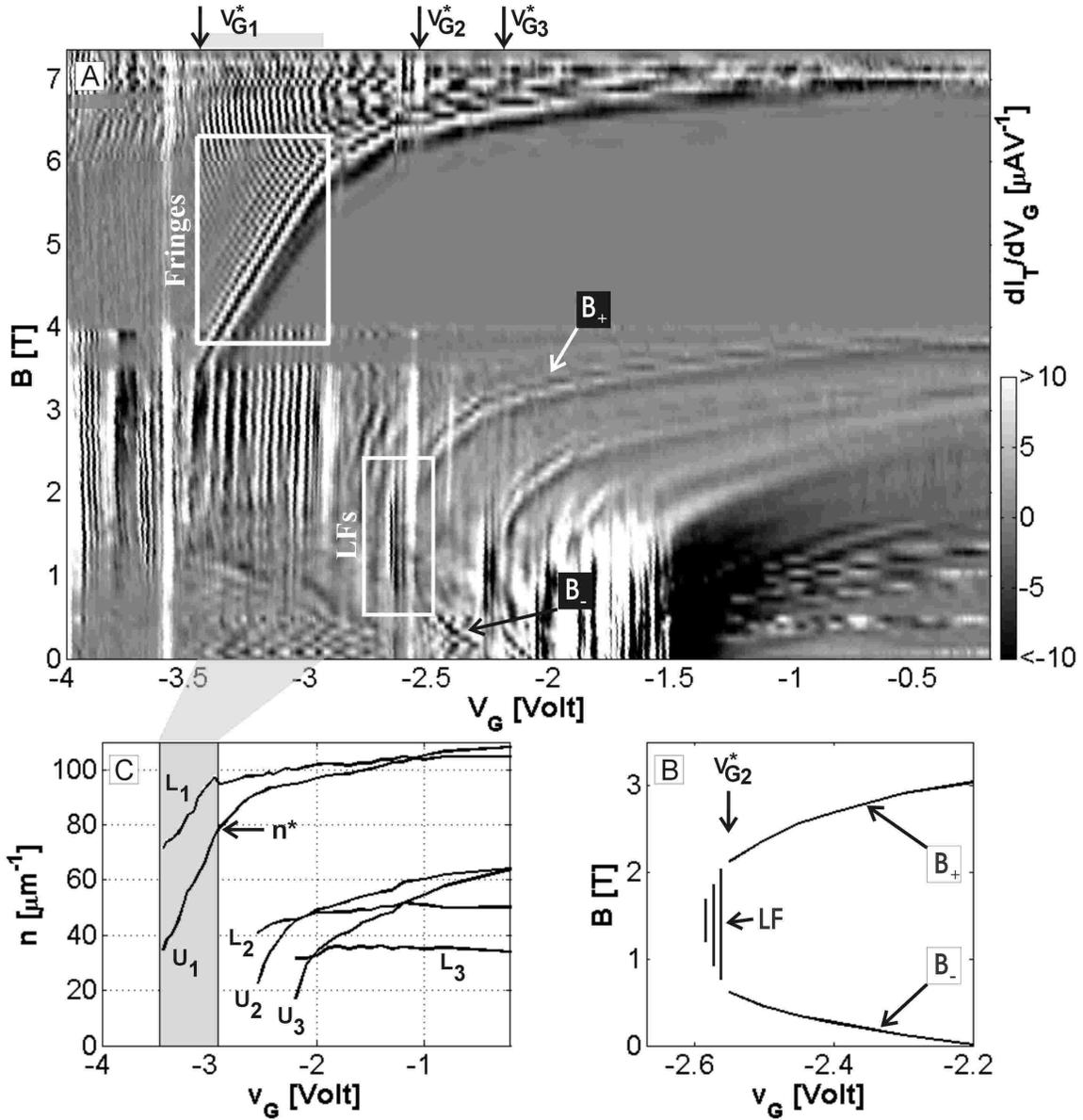}\end{center}
\caption{\label{fig1}{\bf (A)} Gray-scale plot of \dItVgB. {\bf
LFs}: localization features appear for $\vg<{\vg^*}_i$, marked by
arrows; {\bf Fringes}: Finite size fringes. {\bf (B)} Zoom on the
trace of $\bpm$ for sub-band \#2. At ${\vg^*}_2$ $\bpm$ are replaced
by LFs, drawn schematically. {\bf (C)} Dependence of density in
sub-bands \#1-\#3 on \Vg\ ($\uw_i$, $\lw_i$: Upper, Lower wire
sub-band $i$). Gray box marks regime with single occupied sub-band
in each wire ($-3.45V<\vg<-2.90$V), which starts below upper wire
density $n^*$.}
\end{figure}

Interactions become more important as the energy associated with
them increases relative to kinetic energy. To increase this ratio we
reduce electron density in the wires by applying negative voltage,
\Vg, to a 2\micro{m} top gate lying on the surface of the device.
Figure~\ref{fig1}A shows a typical low energy measurement of $\dIt$
as a function of \Vg\ and $B$ \cite{SOM}. The derivative is measured
in order to pick up only the signal from the section of the device
where density is controlled by the gate. This is done by adding a
small ac component to \Vg. A zero-bias anomaly
\cite{Ophir02,Tserkovnyak03} is avoided by setting
$\vsd=100$\micro{V}. This measurement, as well as all those reported
here, is performed at 0.25K.

Figure~\ref{fig1}B shows the typical behavior of $\bpm$. At high
values of \Vg, tunneling is appreciable only in a narrow range
around $\bpm$. As a function of \Vg, $\bpm$ evolve continuously,
following the behavior of $k_{F{\uw}}$ and $k_{F{\lw}}$, allowing us
to invert \eq{Bpm} and extract the density in each sub-band, plotted
in \fig{fig1}C \cite{Ophir02,SOM}. In practice, each wire contains
several sub-bands for most of the \Vg-range. Tunneling is observed
only between sub-bands with the same number of nodes (sub-band
\#1$\rightarrow$\#1, \#2$\rightarrow$\#2 etc.). Tunneling amongst
each pair of sub-bands, one sub-band in each wire, gives rise to a
similar set of features.

\begin{figure}[ptb]
\begin{center}\includegraphics[width=6in]{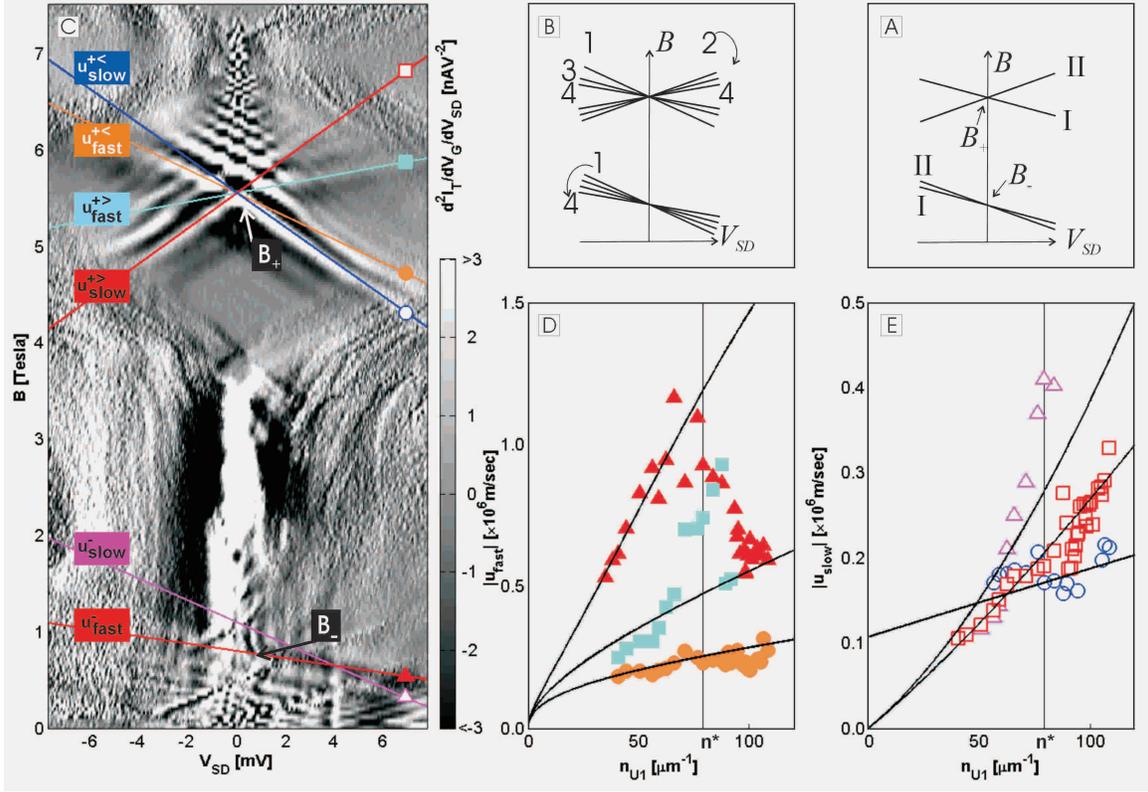}\end{center}
\caption{\label{fig2} {\bf (A) \& (B)} Illustration of the position
of the singularities of \AkE, with $e\vsd$ replacing $E$ and $B$
replacing $k$. {\bf (A)} For non-interacting electrons there are
four curves: $I$ -- copies of lower wire dispersion; $II$ -- copies
of upper wire dispersion. {\bf (B)} With interactions there are ten
curves: three duplicates of each mixed charge mode (marked 3,4), two
copies of each spin mode (marked 1,2). {\bf (C)} Numerical
derivative, with respect to $B$, of the measured \dItVB\ at
$\vg=-3.00$V. Finite size fringes appear for $B>\bp$ and $B<\bm$.
Marked are all extracted slopes, which are offset for clarity.
Triangles-- slopes extracted near \Bm, giving: $\left|u^-_{\tt
fast}\right|$ (filled) $>$ $\left|u^-_{\tt slow}\right|$ (empty),
squares-- positive slopes near \Bp, giving: $u^{+>}_{\tt fast}$
(filled) $>$ $u^{+>}_{\tt slow}$ (empty), circles-- negative slops
near \Bp, giving: $\left|u^{+<}_{\tt fast}\right|$ (filled) $>$
$\left|u^{+<}_{\tt slow}\right|$ (empty). {\bf (D) \& (E)} Apparent
velocities, $u$, versus density, $n_{{\tt U}1}$. Each wire has a
single
occupied sub-band in the shaded region ($n_{{\tt U}1}<n^*$).  %
{\bf (D)} Dependence of $u$'s calculated for small slopes on
density. Overlayed are
curves calculated by setting $v$ to $v_{c-}$ in \eq{u2v} and solving for $u^{\pm}$. %
{\bf (E)} Dependence of $u$'s calculated for large slopes on
density. Overlayed are curves calculated by setting
$v_{s{\ul}}=v_{F{\ul}}/1.25$ in \eq{u2v}.}
\end{figure}

The dispersions of the modes can be determined for every density in
the regime where we observe the \Bpm\ peaks. The dispersions are
traced by the singularities of \AkE\ at finite energy and momentum,
as depicted in \figs{fig2}A,B. For non-interacting electrons,
depicted in \fig{fig2}A, the curves resulting from tunneling either
from or to a Fermi-point, produce a total of four curves: two
shifted copies of the dispersion in each of the two wires
\cite{Ophir02}.

Finite interactions split the singularities of \AkE\ (\fig{fig2}B)
because of two effects. The first is spin-charge separation, caused
by intra-wire interactions, which creates two modes for each
non-interacting mode. The second effect is mode-mixing, caused by
inter-wire interactions. Generally the mixed modes are carried by
both wires, giving rise to four independent velocities \cite{SOM}.
This results in three identical copies of each of the four
dispersions. In the limit of weak tunneling the spin modes do not
couple, and as a result each dispersion branch in \fig{fig2}A splits
into a spin mode and two coupled charge modes, creating four curves
near $\left|q_B\right|=\left|\kfu-\kfl\right|$ and two sets of three
curves near $\left|q_B\right|=\left|\kfu+\kfl\right|$ (\fig{fig2}B).

The tunneling current, $\It(\vsd,B)$, is proportional to \AkE, whose
singularities are peaks for the parameters of our experiment
\cite{Carpentier02}. During a scan of \Vsd\ and $B$, $\It(\vsd,B)$
changes abruptly when the number of modes that can be excited with
the available energy and momentum changes, where \AkE\ is peaked. A
curve along which this happens gives the dispersion of a mode,
$E(k)$. In particular, the slope at $\vsd=0$ gives the dispersion
velocity, $v=\hbar^{-1}\partial E/\partial k$. For the
experimentally relevant case of weak tunneling we expect ten such
intercepts, as shown in \fig{fig2}B, but only four different
magnitudes of slope.

To determine the dependence of the dispersions on density we
measured \dItVB\ for different \Vg's, ranging from $\vg=0$V to
$\vg=-3.45$V. A typical result from the regime where each wire has a
single sub-band, $-3.45<\vg<-2.9$V, is shown in \fig{fig2}C. One can
see that the peaks that appeared in \fig{fig1}A at \Bpm\ split and
move with a slope that gives an apparent velocity:
$u=d^{-1}\left(\partial B/\partial\vsd\right)_{\vsd=0}^{-1}$.
Accompanying these peaks are finite-size fringes
\cite{Tserkovnyak02}.

A total of six slopes appears in \fig{fig2}C, two near \Bm\ and four
near \Bp. Both \Bm\ slopes are negative, giving: $\left|u^-_{\tt
slow}\right|<\left|u^-_{\tt fast}\right|$. Near \Bp\ there are two
negative slopes, giving: $\left|u^{+<}_{\tt
slow}\right|<\left|u^{+<}_{\tt fast}\right|$, and two positive
slopes, giving: $\left|u^{+>}_{\tt slow}\right|<\left|u^{+>}_{\tt
fast}\right|$. For each scan we extracted all discernable slopes.
The results are summarized in \figs{fig2}D,E, where they are plotted
versus the density of electrons in the first sub-band of the upper
wire, $n_{{\uw}1}$. In the shaded area, which extends up to
$n^*\approx80$\micro{m}${}^{-1}$, as extracted from \fig{fig1}A,
only one sub-band is occupied per wire.

\begin{figure}[ptb]
\begin{center}\includegraphics[width=6in]{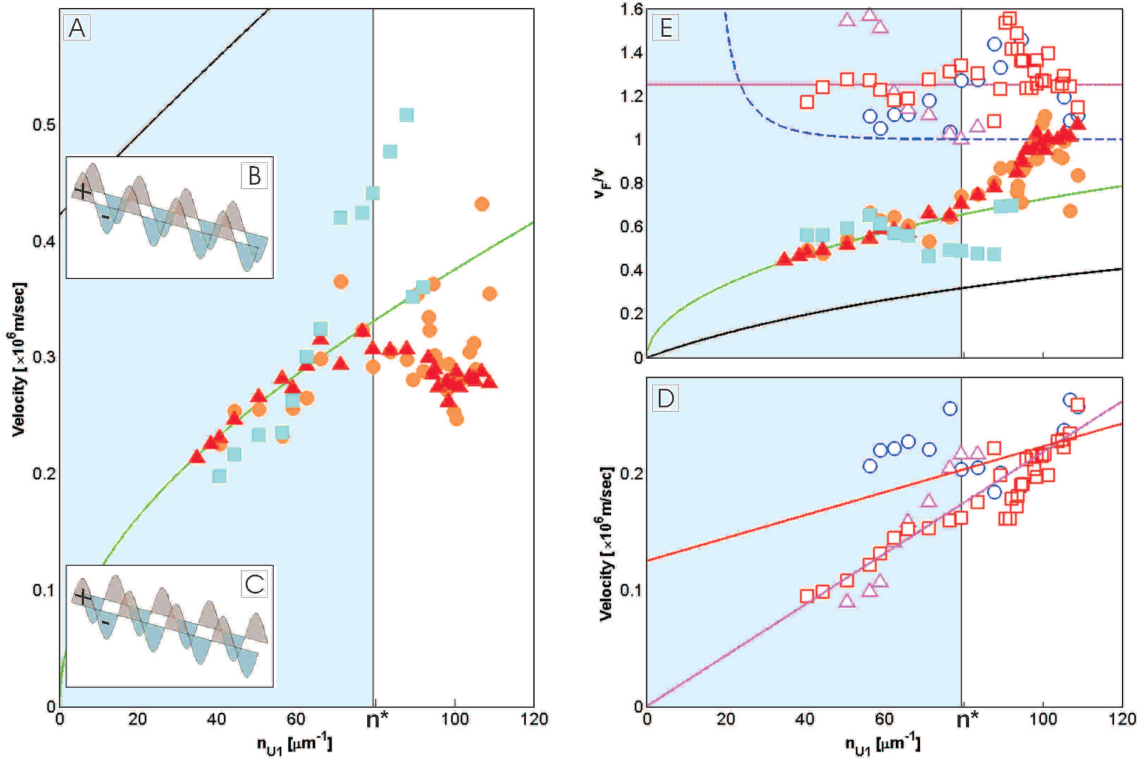}\end{center}
\caption{\label{fig3}Excitation velocity versus density. {\bf (A)}
Velocities obtained from \fig{fig2}D. Curves are the charge
velocities $v_{c-}$ (green) and $v_{c+}$ (black) (see \eq{vmp}).
{\bf (B)/(C)} Illustration of the symmetric / anti-symmetric
coupled-wire mode ($+$: excess positive charge, $-$: excess negative
charge). {\bf (D)} Velocities obtained from \fig{fig2}E. The lines
are $v_{s{\uw}}=v_{F{\uw}}/1.25$ (magenta) and
$v_{s{\lw}}=v_{F{\lw}}/1.25$ (red). The scale is the same as in (A).
{\bf (E)} Plot of $v_F/v$ for the velocities in (A), (B) where $v_F$
is calculated from $n_{{\tt U}1}$ (for $v^-_{\tt fast}$,
$v^{+>}_{\tt fast}$, $v^-_{\tt slow}$, $v^{+>}_{\tt slow}$ \&
$v^{+<}_{\tt fast}$) or from the density in the first sub-band of
the lower wire (for $v^{+<}_{\tt slow}$ and $v_{s{\lw}}$). The
dashed blue line is $f_{\tt pert}$ (see text). The red and magenta
curves from (D) overlap here.}
\end{figure}

The unexpected presence of six different branches of $u$ in
\figs{fig2}D,E wrongly suggests that the coupled-wires have more
than four independent modes. The error lies in assuming that the
band-filling induced by a finite \Vsd\ is negligible \cite{Boese01}.
In reality, \Vsd\ induces charge transfer between the wires, which
is controlled by the mutual capacitance, endowing $\kfu$ and $\kfl$
with a \Vsd-dependence. Thus, the actual excitation velocity is
given by:
\begin{equation}\label{u2v}
    v=\frac{\left|u^\pm\right|}{1\pm\gamma_\pm u^\pm},
\end{equation}
where $\pm$ refers to the crossing point, \Bpm, near which $u^\pm$
is extracted. The value of $\gamma_\pm$ depends on the capacitance
matrix of the wires and is calculated using a simple model
\cite{SOM}. The model consists of two wires of radius $r$, separated
from each other by a distance $d\gg r$ and from a nearby gate by a
distance $D_G/2\gg d$. Since we apply \Vsd\ to the upper wire,
keeping the lower wire grounded, the energetic cost of adding charge
to the wires is given, to quadratic order in excess electron
density, $\delta n_{i}$, by $\sum_{i}\left(E_{Fi}\delta
n_i+e^2c^{-1}_i\delta n^2_i\right)+\frac12\sum_{i,j}e^2\delta
n_ic^{-1}_{ij}\delta n_j-e\delta n_{\uw}\vsd$, where $i,j$ run over
wire indices $\ul$, $E_{Fi}=\hbar^2{k_{Fi}}^2/(2m)$ is the
Fermi-energy, $c^{-1}_i=\pi\hbar/\left(2e^2{v_F}_i\right)$ ($m$ is
the band mass of electrons, ${v_F}_i=\hbar {k_F}_i/m$ is the Fermi
velocity in wire $i$) and $c^{-1}_{ij}$ are elements of the inverse
capacitance matrix. In the random phase approximation
\cite{Nozieres99}, the first term in this expression is kinetic
energy, the second is Coulomb interaction energy. By assumption, the
inverse-capacitance of each wire to the gate ($c^{-1}_{\uw\uw}$ and
$c^{-1}_{\lw\lw}$) is identical:
$c^{-1}_{G}=(2\pi\epsilon)^{-1}\log{[D_G/r]}$. The
inverse-capacitance between the wires
$c^{-1}_{\uw\lw}=c^{-1}_{\lw\uw}$ is:
$c^{-1}_{M}=(4\pi\epsilon)^{-1}$ $\log{\left[1+(D_G/d)^2\right]}$.
Here $r=10$nm and $d=30$nm. $D_G=70$nm is the distance of the wires
to a parallel layer of dopants. Using $D_G=500$nm, the distance to
the top gate, has only minor influence because of the $\log$. The
dimensions are roughly MBE growth parameters and were not adjusted.
The result of applying \eq{u2v} is presented in \fig{fig3}. Clearly
the model is successful when each wire has only one occupied
sub-band: all three fast branches collapse on a single curve for
$n_{{\uw}1}<n^*$ and all slow branches collapse on two curves
\cite{SOM}.

The same model for interactions, that corrects for band-filling,
allows to identify the branches in \fig{fig3}. For this we turn to
the Hamiltonian of the coupled wires, which consists of a free
electron part and an interacting part. Taking a long length
approximation and bosonizing \cite{Matveev93a}:
\begin{equation}\label{hamiltonian}
H=\sum_{i=1}^{N}\sum_{s=\uparrow\downarrow}\int_0^\infty\!\!\!
dx\left[\frac{p_{is}^2(x)}{2mn_{is}}+ \frac{m}2n_{is}
v_{F{is}}^2{q'_{is}}\!^2(x)\right]+\sum_{i,j=1}^{N}\frac{e^2c^{-1}_{ij}}2
\sum_{s,s'=\uparrow\downarrow}n_{is}n_{js'}\int_0^\infty\!\!\!
dxq'_{is}(x)q'_{js'}(x).
\end{equation}
The sums run over all $N$ occupied sub-bands and over both spin
orientations. The density of spin orientation $s$ in sub-band $i$ is
$n_{is}=n_i/2$, $q'_{is}$ is the gradient of the displacement
operator and $p_{is}$ is the conjugate momentum.

Within this model, which neglects back-scattering, the velocities of
the coupled-wire modes are found by diagonalizing \eq{hamiltonian}
using a canonical transformation. This yields spin velocities equal
to the Fermi-velocities. For a single mode in each wire, $N=2$, the
two charge velocities are \cite{Matveev93a}:
\begin{equation}\label{vmp}
{v_c}^2_\pm=\frac{v^2_{c{\uw}}+
v^2_{c{\lw}}}2\pm\left[\left(\frac{v^2_{c{\uw}}-
v^2_{c{\lw}}}2\right)^2+v_{F{\uw}}v_{F{\lw}}
\left(\frac{2e^2}{\pi\hbar}c^{-1}_M\right)^2 \right]^{1/2}\!\!\!\!.
\end{equation}
Here $v_{ci}$ are the charge velocities in each individual wire
\cite{Kane92b}: $v_{ci}/v_{Fi}=\sqrt{1+U_i/(2E_{Fi})}$, where
$U_i=e^2n_i/c_G$ is the interaction energy. Physically $+/-$
correspond to symmetric / antisymmetric excitations (illustrated in
\figs{fig3}B,C). When the wires are identical both modes are carried
equally by both wires, but when the densities differ, as in the
experiment, the symmetric mode is carried primarily by the more
occupied wire, the lower wire, while the antisymmetric mode is
carried primarily by the upper wire.

We have overlaid the result of \eq{vmp} on the corrected velocities
in \figs{fig3}A,E. The fast velocities follow the calculated
${v_c}_-$ closely for $n_{{\uw}1}<n^*$, attesting to the validity of
the model and leading us to associate them with the antisymmetric
charge mode. The faster ${v_c}_+$, on the other hand, is completely
absent from the data. This is to be expected near \Bm, where
tunneling creates interacting electron-hole pairs which propagate
together, excitations that are almost completely anti-symmetric. On
the other hand, near \Bp\ none of the excitations branches should be
suppressed, as they are all excited by tunneling, leaving the issue
of the complete absence of the symmetric mode unresolved.

Turning to the slow branches in \figs{fig3}D,E, we find linear
dependence on the bare Fermi-velocities, $v_{si}=v_{Fi}/f_i$, where
$f_{\uw}=f_{\lw}=1.25$. The linearity and the fact that $f_{i}>1$,
suggest that these modes are the spin modes. Theoretically one
expects a spin velocity equal to $v_F$ as long as back-scattering is
small, while finite back-scattering reduces it below $v_F$
\cite{Coll74,Schulz93,Creffield01,Hausler02,Matveev04,Cheianov04}.
For example, one group \cite{Creffield01,Hausler02}, using Monte
Carlo simulations, finds that the expression $f_{\tt
pert}=1/\sqrt{1-V_{2k_F}/\left(\pi\hbar v_F\right)}$ \cite{SOM},
gives an upper bound to the ratio between the $v_F$ and $v_s$.
However, a plot of $f_{\tt pert}$ in \fig{fig3}E shows that $f_{\tt
pert}$ does not account for the deviation of $f_{i}$ from unity.

Further examination of \fig{fig1}A down to the depletion of each
sub-band in the upper wire reveals that the continuous evolution of
each set of \Bpm\ peaks is replaced by a series of vertical streaks,
dubbed localization features (LFs). These occur in a small range of
\Vg\ below ${\vg}_i^*=-3.45,~-2.55,~-2.20$V, for each of the three
upper-wire sub-bands we observe. Each set of LFs signals an abrupt
change in the momentum-space content of the wavefunction in the
depleting sub-band.

Above ${\vg}_i^*$ finite-size fringes for $B<\bm$ and $B>\bp$
accompany the $\bpm$-peaks, signifying that the states in both wires
contain only wavenumbers higher than the Fermi wavenumber and
implying that they are extended \cite{Tserkovnyak02}. The location
of the fringes at $B<\bm$ and $B>\bp$ indicates that the potential
along the non-uniform upper wire has a hump, with a typical length
given by the period: $h/\left(e\Delta B_{\rm
fringe}d\right)\approx0.75$\micro{m}, consistent with the barrier
the surface gate induces.

Below ${\vg}_i^*$, we find that each LF fills a broad range in $B$,
lying roughly between the extrapolations of $\bpm$. This implies
that the wavefunction of the state along the upper wire is
localized. We are thus led to conclude that the $B$-streaks signify
a qualitative change in the self-consistent potential at
${\vg}_i^*$, which marks a transition between an extended state and
a localized state. The localized states appear while the more
occupied sub-bands are still fully conducting, in contrast to a
recent study of inhomogeneous wires \cite{Thomas04}.

\begin{figure}[ptb]
\begin{center}\includegraphics[width=4in]{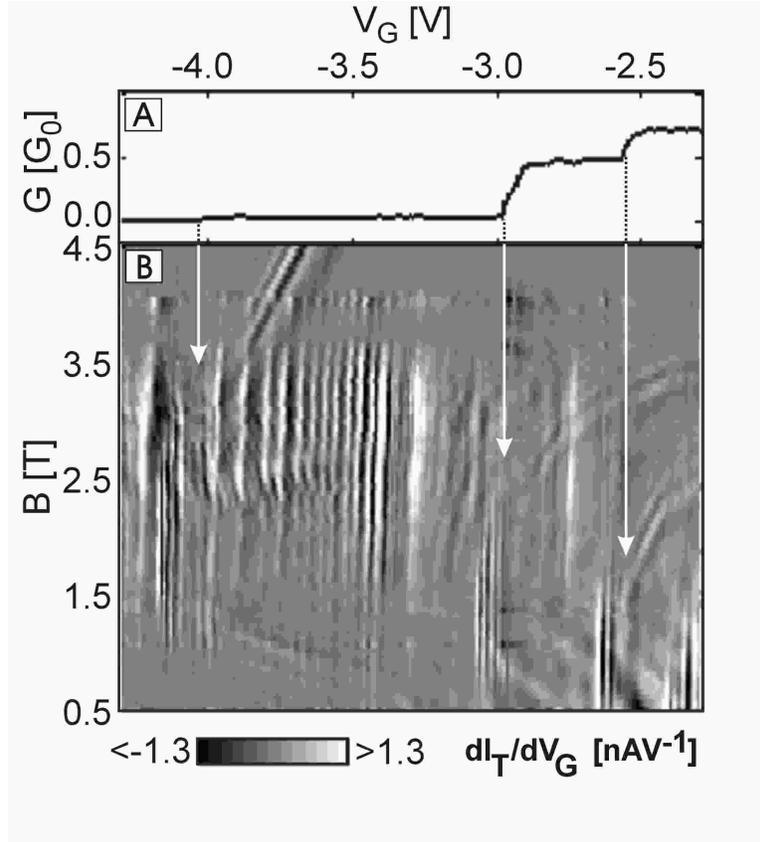}\end{center}
\caption{\label{fig4}{\bf (A)} Two-terminal conductance of the upper
wire as a function of \Vg, measured by applying a dc-voltage bias of
$100\mu$V along the wire. The step height deviates from the
universal $2e^2/h$ because of the indirect upper wire contact
\cite{Yacoby96,dePicciotto01}. This measurement depends very weakly
on $B$, which is $1.3$T here. {\bf (B)} Simultaneous measurement
\dItVgB, showing that localization is concurrent with the
conductance drops.}
\end{figure}

The localization transition affects transport along the upper wire.
Figure~\ref{fig4}A shows the two-terminal conductance along this
wire, $G(\vg,B)$, which is quantized. The stepwise decrease of
$G(\vg,B)$ with density is a hallmark of ballistic transport in a
wire \cite{Beenakker91}. We were able to measure $G(\vg,B)$
simultaneously with \dItVgB, by recording both the dc-current along
the upper wire and the ac-component of the tunneling
current\cite{SOM}. The positions of the steps in $G(\vg,B)$, whose
dependance on $B$ is very weak, are concurrent with the localization
transitions apparent in \fig{fig4}B. We thus conclude that electrons
in a sub-band cease to conduct because of localization while their
density is still finite.

The localization transition hints that bound states come into
existence over the barrier induced by the top gate, which we use to
vary the density. The possibility of this occurring has been
addressed in the context of the 0.7-anomaly in point contacts
\cite{Thomas96,Cronenwett02,Reilly02,dePicciotto04}. Using a variety
of theoretical tools it was found that, when the density is low
enough, a bound state may exist over the barrier
\cite{Meir02,Hirose03,Sushkov03,Matveev04}. Our measurements show
clear evidence for this scenario in long 1D channels. Finally,
0.7-anomaly-like features are observed regularly in the conductance
steps of CEO wires similar to ours \cite{dePicciotto04}. Further
work is needed to show a direct link between the $0.7$-anomaly and
the observed localization features.

\newcommand{\noopsort}[1]{} \newcommand{\printfirst}[2]{#1}
  \newcommand{\singleletter}[1]{#1} \newcommand{\switchargs}[2]{#2#1}

\begin{scilastnote}
\item With pleasure we acknowledge numerous discussions with Greg Fiete, Yuval Oreg  and Jiang Qian.
This work was supported in part by the US-Israel BSF, the European
Commission RTN Network Contract No. HPRN-CT-2000-00125 and NSF Grant
DMR 02-33773. YT is supported by the Harvard Society of Fellows.
\end{scilastnote}

\clearpage

\clearpage \setcounter{equation}{0}%
\setcounter{figure}{0}
\def\thefigure{S\arabic{figure}}
\def\theequation{S\arabic{equation}}
\section*{Supporting text}
The actual values of the charge velocity, $v_c$, and the spin
velocity, $v_s$, depend on microscopic details and are very
difficult to determine, both theoretically and in experiment. Of
particular interest is their dependence on $n$, which directly
controls the ratio between the Coulomb interaction and kinetic
energy. A one-dimensional charge mode, which resembles a charge
density wave, travels with a velocity that is strongly affected by
the Coulomb interaction: $v_c=v_F/g$. Here the Fermi-velocity is a
measure of electron density and $g$ is a measure the relative
strength of the Coulomb interaction. For repulsive interactions
$0<g<1$, while for non-interacting electrons $g=1$. Within the
random phase approximation, which gives reliable estimates for $g$
if backscattering is weak, it is found that as $n$ is reduced, $g$
decreases.

The propagation velocity of the spin modes, $v_s$, is related to
exchange interaction. According to theory, $v_s$ is suppressed for
very strong repulsive interactions, where it is difficult for
adjacent electrons to exchange places, leading to $v_s\ll v_F$
\cite{Schulz93s,Hausler02s,Matveev04s,Cheianov04s}. The main text
cites an example for the suppression factor of $v_s$ relative to
$v_F$ in one model \cite{Hausler02s,Creffield01s}, which is given by
$1/f_{\tt pert}=\sqrt{1-V_{2k_F}/\left(\pi\hbar v_F\right)}$. Here
$V_k$ is the Fourier transform of the two-body interaction potential
in a single wire in the presence of a gate. It is given by:
$$%
\frac{V_{2k_F}}{\pi\hbar
v_F}=\frac{2}{a_Bk_F}\left[K_0\left(2kr\right)-
K_0\left(2k\sqrt{r^2+{D_G}^{2}}\right)\right], %
$$%
where $K_0$ is a Bessel function, $a_B$ is the Bohr radius in GaAs
and $D_G$ is the distance to the top gate, $500$nm.

A remark is due on the number of expected singularity branches
crossing the $\vsd=0$ axis in a measurement of \dItVB. In principle,
even in the absence of inter-wire interactions, we expect for each
charge mode an extra feature with opposite slope and very small
amplitude (unless interactions are very strong). This gives an extra
copy of each charge mode branch near \Bm. The reason is that forward
scattering has a contribution from interactions between left and
right movers, causing an electron tunneling into one branch of
movers to induce a density modulation in the other branch as well.

\begin{figure}[ptb]
\begin{center}\includegraphics[width=4in]{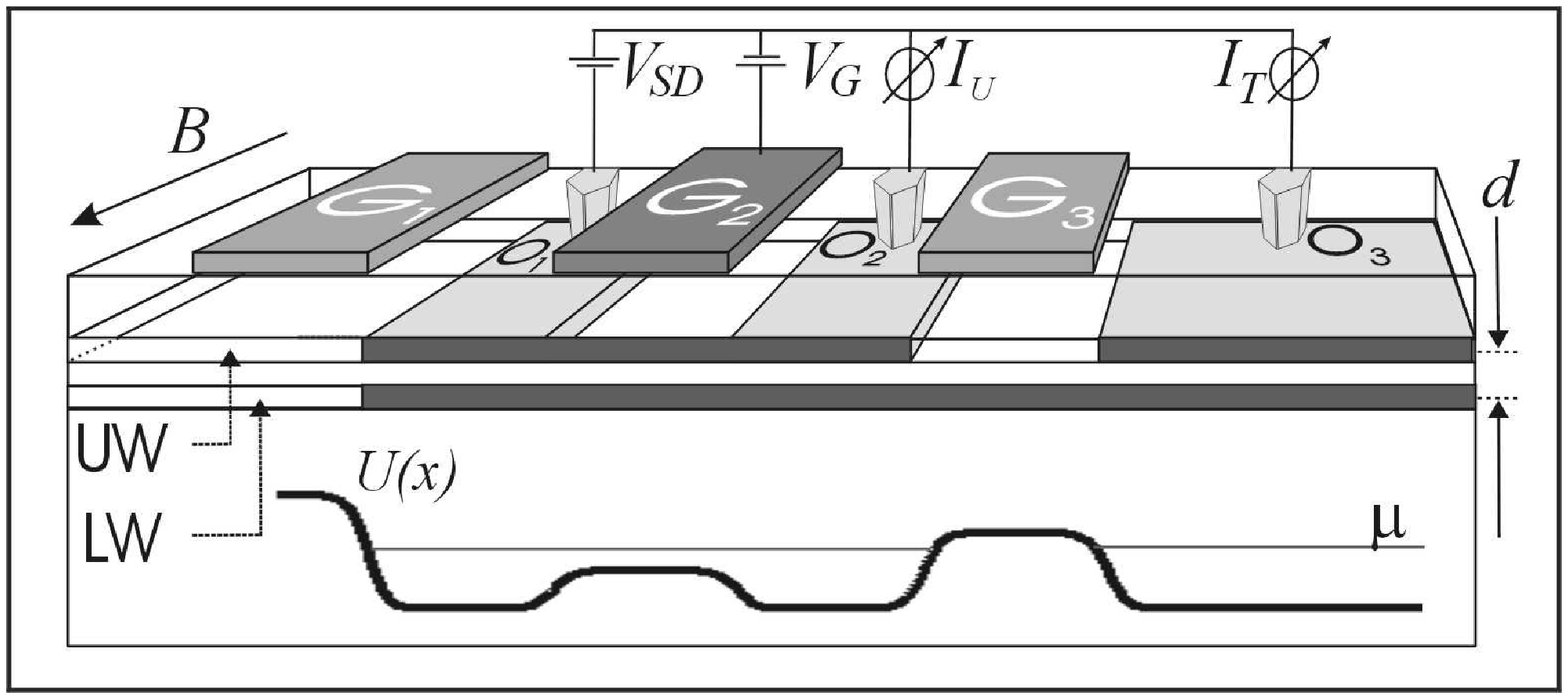}\end{center}
\caption{\label{fig1S}%
Schematic of the device, cleavage-plane facing front, perpendicular
to $B$. Depicted: 2\micro{m}-wide top gates (G\sub{1-3}), ohmic
contacts to 2DEG (O\sub{1-3}), 20nm-wide upper wire at edge of 2DEG,
30nm-wide lower wire and 6nm insulating AlGaAs barrier, \Vg: voltage
applied to control density, $I_U$: current through upper wire,
$\It$: tunneling current. Also illustrated: $U(x)$, gate-induced
potential, and electrochemical potential in the upper wire, $\mu$.}
\end{figure}
Our device consists of two parallel wires separated by a tunnel
junction at the edges of two quantum wells in a GaAs/AlGaAs
heterostructure created by cleaved edge overgrowth \cite{Ophir02s}.
The band structure is such that besides the wires along the edge,
the upper well is also occupied by a two-dimensional electron gas
(2DEG), which we use to contact the wires. The measurements were
conducted in a ${}^3$He-fridge at a base temperature of $0.25$K.
After the sample cools down, it is illuminated by infra-red light,
which ionizes impurities and increases the overall electron density
in the device.

Tungsten top-gates, 2\micro{m}-wide, lying $500$nm above the tunnel
junction and 2\micro{m} apart, control the density in sections of
the device. Both the tunnel junction and the upper wire are
delimited by applying voltage to two peripheral gates, G\sub1 \&
G\sub3 in \fig{fig1S}, lying 10\micro{m} apart. Bias voltage \Vg\ is
applied to a central gate, G\sub2 in \fig{fig1S}, to vary the
density in the central section of the device. To increase
sensitivity to processes affected by G\sub2, we measure the
derivative of $\It$ with respect to \Vg. To this end we add a small
ac-component to \Vg\ (a few mV at a few Hz), apply finite \Vsd\ to
contact O\sub1 and pick up the resulting current at contact O\sub3
with a lock-in amplifier. An additional contact, O\sub2, can be
grounded in order to measure the current along the upper wire,
$I_U$, but is left floating otherwise. To measure $G(\vg,B)$ we
apply voltage $V$ to contact O\sub1 (cf. \fig{fig1S}) and ground
contact O\sub2. This gives net bias $\vsd=\alpha V$ relative to the
lower wire (where $0<\alpha<1$) and a bias drop of $V$ along the
upper wire.

In the \Vg-range being studied we do not observe transitions between
a sub-band in one wire and more than one sub-band in the other wire.
This hints at a selection rule, which we conjecture arises from the
similarity of the wavefunctions in the two wires in the planes
perpendicular to them \cite{Tserkovnyak03s}. According to the rule,
the overlap of wavefunctions in different wires with a different
number of nodes perpendicular to the wires is small and suppresses
the transitions between them. In identical wires this selection rule
would be absolute, because wavefunctions from different sub-bands
would be orthogonal.

The finite width of the wires in the direction perpendicular to $B$
distorts the dispersions slightly. Equation~1 in the main text is
precise only as long as a single sub-band is occupied in each wire.
When higher sub-bands are occupied, finite $B$ flattens the
dispersions \cite{Berggren86s}, changing the occupations and
generally making it energetically favorable to occupy lower
sub-bands at the expense of higher ones. We ignore this distortion
as it gives an error that is comparable to the accuracy of the
measurement.

In the main text we describe a model that allows to calculate the
actual excitation velocities from the measured slopes. According to
the model,  the value of $\gamma_\pm$ appearing in Eq.~2 in the main
text is given by:
\begin{equation}\label{gamma}
\gamma_\pm=\frac{\pi\hbar}{e^2}\left[c_{\pm}
c_{\uw}\left(c_{\lw}+c_\mp/2\right)\right]\left[\left(c_++c_-\right)
\left(c_{\lw}+c_{\uw}\right)+c_{+}c_{-}+4c_{\lw}c_{\uw}\right]^{-1},
\end{equation}
where $c^{-1}_{\pm}=\left(c^{-1}_{G}\pm c^{-1}_{M}\right)/2$, while
the rest of the quantities here are given in the main text.

The model has a limited range of validity. Above $n^*$, as more
sub-bands are occupied, it exaggerates the voltage-induced
band-filling. In this regime, the sparsely occupied sub-bands are
filled instead of the more populated ones, because they are more
compressible. Thus in this regime Eq.~2 corrects the $u$'s too much,
bringing the corrected velocities too low, as can be seen for
$v^{+<}_{\rm fast}$, $v^{-}_{\rm fast}$ in Fig.~3. To obtain the
fits shown in this figure we used the same values of $r$ and $d$ as
for the band-filling correction. For $D_G$ we used here $500$nm, the
distance to the metallic top gate, rather than the distance to the
dopant layer, because typical timescales for the reaction of that
layer are too slow to influence the dynamics of the modes.

\renewcommand{\noopsort}[1]{} \renewcommand{\printfirst}[2]{#1}
  \renewcommand{\singleletter}[1]{#1} \renewcommand{\switchargs}[2]{#2#1}

\end{document}